\documentclass[conference]{IEEEtran}
\IEEEoverridecommandlockouts
\usepackage{cite}
\usepackage{amsmath,amssymb,amsfonts}
\usepackage{algorithmic}
\usepackage{graphicx}
\usepackage{textcomp}
\usepackage{xcolor}
\usepackage{physics}
\usepackage{caption}
\usepackage{subcaption}
\def\BibTeX{{\rm B\kern-.05em{\sc i\kern-.025em b}\kern-.08em
    T\kern-.1667em\lower.7ex\hbox{E}\kern-.125emX}}

\begin{document}

\title{Charged Particle Motion in Neutron Star Magnetic Fields: A Comparison Between the Boris Algorithm and the Guiding Center Approximation}

\author{\IEEEauthorblockN{Minghao Zou}
\IEEEauthorblockA{
\textit{Valley Christian High School}\\
Santa Clara, USA \\
minghao2006.zou@gmail.com}
\and
% \IEEEauthorblockN{Bart Ripperda}
% \IEEEauthorblockA{\textit{School of Natural Sciences} \\
% \textit{Institute for Advanced Study}\\
% 1 Einstein Drive, Princeton, NJ 08540, USA \\
% bartripperda@ias.edu}
\IEEEauthorblockN{Bart Ripperda}
\IEEEauthorblockA{\textit{School of Natural Sciences} \\
\textit{Institute for Advanced Study}\\
New Jersey, USA \\
bartripperda@ias.edu}
}

\maketitle
%979-8-3503-0965-2/23/$31.00
\IEEEoverridecommandlockouts
\IEEEpubid{\makebox[\columnwidth]{ 979-8-3503-0965-2/23/\$31.00~\copyright~2023~IEEE \hfill}
\hspace{\columnsep}\makebox[\columnwidth]{ }}

\begin{abstract}
% Emission from neutron stars originates typically from their magnetosphere due to radiating electrons and positrons. The motion of these relativistic charged particles in the strong and complex magnetic fields is governed by the Lorentz force. Relativistic charged particle trajectories under uniform magnetic and electric fields can be easily calculated analytically. However, under more complex fields such as magnetic dipolar or quadrudipolar fields, numerical solutions are required. Two often employed schemes are the Boris method to calculate the solution to the full equations of motion, and the guiding center approximation (GCA), which can reduce computational complexity when the Larmor radius is sufficiently small, by evolving the orbital center rather than the particle itself. We compare both methods to calculate particle motion in a series of basic tests, discuss their characteristics and quantify their accuracy. We proceed to apply the methods to dipolar, quadrupolar, and quadrudipolar magnetic fields, as applicable for magnetospheres. It is essential to consider such realistic magnetic field geometries around neutron stars to model the emission from magnetars and pulsars. Results of the simulation can assist the Neutron Star Interior Composition ExploreR (NICER) in understanding emission from nearby the stellar surface and its magnetosphere and to study its composition.
Neutron star emission originates typically from its magnetosphere due to radiating electrons. Trajectories of relativistic charged particles under uniform electromagnetic fields can be calculated analytically. However, under more complex and realistic fields, numerical solutions are required. Two common schemes are the Boris method, which solves the full equations of motion, and the guiding center approximation (GCA), which only evolves the orbital center. We compare both methods in a series of tests, discuss their characteristics and quantify their accuracy. We apply the methods to dipolar, quadrupolar, and quadrudipolar magnetic fields, as applicable for magnetospheres. It is essential to consider such realistic neutron star magnetic field geometries to model the emission from magnetars and pulsars. Our work can assist the Neutron Star Interior Composition ExploreR (NICER) to understand emission from neutron stars’ surface and magnetosphere and to study its composition.
\end{abstract}

\section{Introduction}
%lets cite following here:
%https://arxiv.org/pdf/2209.11362.pdf
%https://arxiv.org/pdf/1604.04625.pdf
%https://arxiv.org/pdf/1510.01734.pdf
%https://arxiv.org/pdf/2001.02236.pdf
%https://arxiv.org/pdf/1704.05062.pdf`
Particle simulations are essential to study the typically collisionless, relativistic plasma flow near black holes and neutron stars. Conventional magnetohydrodynamics models are useful to capture the large-scale motion of the plasma, while the kinetic equations of motion describing the evolution of particle distributions is essential to resolve small-scale plasma effects in the strong dynamic magnetic field.

Particle movement in uniform electric and magnetic fields can be calculated easily analytically. For example, a non-relativistic particle with velocity $v_\perp$ perpendicular to a uniform magnetic field gyrates with respect to some center with Larmor radius (i.e. gyro-radius) $R_c = mv_\perp/qB$. Adding a parallel component to the velocity $\textbf{v}_\parallel = v_\parallel \hat{\textbf{z}}$ results in a helix defined by $(R_c\cos(t), R_c\sin(t), t)$ for a parametric variable $t$ relating to time. In more realistic and complex fields (e.g. magnetic dipole or higher order poles and combinations thereof), analytical calculations become much more difficult. Thus, discrete numeric methods are required, where we calculate the particle position and momentum with a small time step $\Delta t$ and update the relevant parameters to calculate the next position. 

In this paper we examine two popular methods to solve the equations of motion for charged particles: the Boris scheme and guiding center scheme described in \cite{Ripperda2018}, \cite{Bacchini2020}, and \cite{mignone2023}. 
\vfill\null
While the Boris pusher solves the full equations of motion, the GCA only resolves the trajectory for the center of gyration of the particle while updating $v_\perp$, the component of velocity perpendicular to the magnetic field (and thus no information on the trajectory is lost, except the angle of gyration). We will perform a series of tests with these two methods, namely, the often occurring $\mathbf{E} \times \mathbf{B}$ drift and motion in and around a magnetic null (e.g. typical around a current sheet).

We will also investigate particle motion in realistic fields for neutron star magnetospheres, i.e., magnetic dipolar, quadrupolar, and quadrudipolar fields, the latter two of which have been given only theoretical treatments in the context of neutron stars in \cite{Gralla2016} and \cite{Gralla2017}. Based on (general relativistic) particle-in-cell simulations, simultaneously solving Maxwell's equations and the particle equations of motion for dipolar magnetic fields, it is known that particles can be produced through quantum electrodynamics processes nearby the polar cap of the neutron star \cite{Philippov2015} and in the separatrix between the last closed field line of the dipole and the first open field line \cite{Bransgrove2022}. These particles can fall onto and bombard the star, potentially resulting in X-ray emission. In these regions, strong discharges due to pair creation and acceleration can also result in coherent radio emission from nearby the pulsar \cite{Philippov2015} \cite{Bransgrove2022}. To better understand and model the emission, it is thus important to consider more realistic magnetic field geometries. While a dipolar magnetic field serves as a good and often employed model, observations made by the Neutron Star Interior Composition ExploreR (NICER) on the International Space Station suggest that neutron star magnetic fields may include quadrupole and quadrudipolar components. Thus, understanding particle motion in these field geometries can help us uncover a more accurate depiction of plasma flow near neutron stars and its emission. To resolve the radiation mechanism, i.e., an electron moving in a magnetic field emitting or scattering a photon, it is essential to accurately capture the trajectory of the particle in the highly magnetized plasma. However, at high magnetization and thus small Larmor radii, the Boris method becomes very expensive to accurately capture the gyration; therefore we consider applying the GCA, which realistically captures the trajectory of a particle in a strong magnetic field without the requirement to resolve the small Larmor radius. Both the Boris method and GCA will thus be tested for a range of electromagnetic fields, and their accuracy will be discussed. In this paper we will use SI units and normalize $c = 1$.

\section{Numerical Methods}
A particle with charge q, mass m, and velocity vector $\mathbf{v}$, in an arbitrary electric field \textbf{E} and magnetic field \textbf{B} undergoes motion described by \begin{equation}\frac{d\textbf{u}}{dt} = \frac{\textbf{F}}{m} = \frac{q}{m}(\textbf{E} + \textbf{v}\times \textbf{B}),\label{eq:1}\end{equation} where $\textbf{F}$ is the force experienced by the particle and $\textbf{u}$ is its relativistic momentum over mass. Discretizing this equation yields \cite{Ripperda2018},
\begin{gather}
    \textbf{x}^{n + 1/2} = \textbf{x}^n + \frac{\textbf{u}^n}{2\gamma^n} \Delta t, \label{eq:2}\\
    \frac{\textbf{u}^{n+1}-\textbf{u}^{n}}{\Delta t} = \frac{q}{m} \left[ \textbf{E}(\textbf{x}^{n+1/2}) + \bar{\textbf{v}} \times \textbf{B}(\textbf{x}^{n+1/2})\right], \label{eq:3}\\
    \textbf{x}^{n+1} = \textbf{x}^{n+1/2} + \frac{\textbf{u}^{n+1}}{2\gamma^{n+1}} \Delta t. \label{eq:4}
\end{gather}
Here, the Lorentz factor is $\gamma = 1/\sqrt{1-v^2}$, $v$ is the magnitude of the velocity vector, and $\bar{\textbf{v}}$ represents the average velocity between timesteps $n$ and $n+1$.  However, the implicit nature (i.e. the need to know information from the next time step) of $\bar{\textbf{v}}$ and the non-linearity of the $\gamma$ factor necessitates the introduction of approximations to develop a numerical scheme.

\subsection{Boris Method}
The Boris scheme is one of the most popular methods used in relativistic particle simulation, primarily due to its ability to closely track particle trajectories at high Lorentz factors and its energy-conservation nature. However, the Boris method is prone to develop a phase delay in the gyromotion. 

The Boris scheme defines $\bar{\textbf{v}}$ as \begin{gather}\bar{\textbf{v}} = \frac{\textbf{u}^{n+1} + \textbf{u}^n}{2\gamma^{n+1/2}}.\label{eq:5}\end{gather} Substituting this definition into equation \ref{eq:3} results in the following \cite{Birdsall&Langdon}:
\begin{gather}
    \textbf{u}^- = \textbf{u}^n + k\textbf{E}^{n+1/2}, \label{eq:6}\\
    \textbf{u}' = \textbf{u}^- + \textbf{u}^- \times \textbf{t}, \label{eq:7}\\
    \textbf{u}^+ = \textbf{u}^- + \frac{2}{1+t^2} \textbf{u}' \times \textbf{t}, \label{eq:8}\\
    \textbf{u}^{n+1} = \textbf{u}^+ + k\textbf{E}^{n+1/2}. \label{eq:9}
\end{gather}
Here, $k = {q \Delta t}/{2m}$, $\textbf{t} = k\textbf{B}^{n+1/2}/\bar{\gamma}^{n+1/2}$, and we approximate $\bar{\gamma}^{n+1/2} \equiv \sqrt{1+(\textbf{u}^-)^2}$.

\subsection{Guiding Center Method}

The GCA is useful when the length scale of the local $\grad B$ is much greater than the local Larmor radius \cite{Ripperda2018}. This can occur e.g., due to the rapid decrease of the Larmor radius of a strongly radiating particle to one so small that the original Boris time step cannot resolve the particle's gyrofrequency, resulting in inaccurate trajectories. 

In the GCA, however, only the gyro center position and the velocity perpendicular to the magnetic field of the particle is updated. Due to the conservation of magnetic moment, information on the particle orbital is thus preserved while only calculating the orbital center. In the following equations, we choose to discard mirror effects as, for the purpose of simulations in compact-object magnetospheres, typically $v_\perp \ll 1$, where $v_\perp$ is the particle velocity perpendicular to the magnetic field \cite{Philippov2020}. The time derivative of the guiding center position and momentum are given by \cite{Bacchini2020},
\begin{equation}\label{eq: 10}
    \begin{gathered}
        % \dot{\textbf{R}} = v_\parallel \textbf{b} - \frac{\textbf{b}\times c\textbf{E}}{B} + \frac{\textbf{b}}{B\left(1-\frac{E_\perp^2}{B^2}\right)}  \times \Bigg[\Bigg. \frac{cm\gamma}{q} \\ \biggr[\biggr. v_\parallel^2 (\textbf{b} \cdot \grad) \textbf{b} + v_\parallel (\textbf{v}_E \cdot \grad) + \\ v_\parallel (\textbf{b} \cdot \grad) \textbf{v}_E + (\textbf{v}_E \cdot \grad ) \textbf{v}_E \biggr.\biggr] + \\ \frac{\mu c}{\gamma q} \grad \left[ B\left( 1-\frac{E_\perp^2}{B^2}\right)^{1/2}\right] + \frac{v_\parallel E_\parallel}{c} \textbf{v}_E \Bigg. \Bigg],
        \dot{\textbf{R}} = \frac{\textbf{u}_\parallel}{\Gamma} + \textbf{v}_E + \textbf{c},
    \end{gathered}
\end{equation}
\begin{equation}\label{eq:11}
    \begin{gathered}
        % \dot{u}_\parallel = \gamma\textbf{v}_E \cdot \left(v_\parallel(\textbf{b}\cdot\grad)\textbf{b} + (\textbf{v}_E \cdot \grad)\textbf{b}\right) + \frac{qE_\parallel}{m} - \\ \frac{\mu}{\gamma m}\textbf{b}\cdot\grad\left[B\left(1-\frac{E_\perp^2}{B^2}\right)^{1/2}\right].
        \dot{u}_\parallel = \frac{q}{m}E_\parallel + u_\parallel \textbf{v}_E \cdot (\textbf{b}\cdot \grad)\textbf{b} + \Gamma \textbf{v}_E \cdot (\textbf{v}_E \cdot \grad) \textbf{b}.
    \end{gathered}
\end{equation}
Here, $\textbf{R}$ is the position vector of the guiding center. $\textbf{b}$ is the unit vector in the direction of $\textbf{B}$. $E_\perp$ is the electric field component perpendicular to $\textbf{B}$ and $\textbf{E} = \textbf{E}_\perp + \textbf{E}_\parallel.$ $\textbf{v}_E = \textbf{E}\times \textbf{b}/B$ is the $\textbf{E}\cross \textbf{B}$ drift velocity vector and $\textbf{c}$ is the curvature drift velocity vector, which is defined as,
\begin{equation}\label{eq:12}
    \begin{gathered}
    \textbf{c} = \frac{m\kappa^2}{qB}\textbf{b}\times \left[\frac{u_\parallel^2}{\Gamma}(\textbf{b}\cdot\grad)\textbf{b} + u_\parallel (\textbf{v}_E\cdot\grad)\textbf{b}\right].
    \end{gathered}
\end{equation}

Here, $\kappa = 1/\sqrt{1-v_E^2}$ is the Lorentz factor associated with $v_E$. Discretizing these equations yields the following \cite{Bacchini2020}:
\begin{equation}\label{eq:13}
    \begin{gathered}
    \frac{\textbf{R}^{n+1}-\textbf{R}^n}{\Delta t} = \frac{u^{n+1/2}_\parallel}{2}\left(\frac{\textbf{b}^{n}}{\Gamma^n_{n+1/2}} + \frac{\textbf{b}^{n+1}}{\Gamma^{n+1}_{n+1/2}}\right) + \\ \frac{\textbf{v}_E^n + \textbf{v}_E^{n+1}}{2} + \frac{\textbf{c}^{n}_{n+1/2} + \textbf{c}^{n+1}_{n+1/2}}{2},
    \end{gathered}
\end{equation}
\begin{equation}\label{eq:14}
    \begin{gathered}
    \frac{u_\parallel^{n+\frac{1}{2}} - u_\parallel^{n-\frac{1}{2}}}{\Delta t} = \frac{q}{m} E_\parallel^n + \frac{u_\parallel^{n+\frac{1}{2}} + u_\parallel^{n-\frac{1}{2}}}{2}\textbf{v}_E^n \cdot \\ (\textbf{b}^n \cdot \grad)\textbf{b}^n + \frac{1}{2}\left(\Gamma^n_{n-1/2} + \Gamma^n_{n+1/2}\right)\textbf{v}_E^n \cdot (\textbf{v}_E^n\cdot\grad)\textbf{b}^n,
    \end{gathered}
\end{equation}
Here, $\parallel$ refers to component of a vector in the direction parallel to the magnetic field. $\Gamma = \sqrt{1+u^2}$ is the Lorentz factor associated with $u$.  Any auxiliary variable $A^{n_1}_{n_2}$ is defined as $A(\textbf{R}^{n_1}, u_\parallel^{n_2})$. 

Solving equation \ref{eq:14} for $u_\parallel^{n+1/2}$ yields,
\begin{equation}\label{eq:15}
    \begin{gathered}
    u_\parallel^{n+1/2} = \left[u_\parallel' + \frac{\Delta t}{2}\Gamma^n_{n+1/2}\textbf{v}_E^n\cdot(\textbf{v}_E^n\cdot\grad)\textbf{b}^{n}\right]\cdot \\ \left[1-\frac{\Delta t}{2}\textbf{v}_E^n\cdot(\textbf{b}^n\cdot\grad)\textbf{b}^n \right] ^{-1},
    \end{gathered}
\end{equation}
where
\begin{equation}\label{eq:16}
    \begin{gathered}
    u_\parallel' = u_\parallel^{n-1/2}\left[1+\frac{\Delta t}{2}\textbf{v}_E^n\cdot(\textbf{b}^n\cdot\grad)\textbf{b}^n\right]+\frac{q \Delta t}{m}E_\parallel^n + \\ \frac{\Delta t}{2}\Gamma^n_{n-1/2}\textbf{v}_E^n\cdot(\textbf{v}_E^n\cdot\grad)\textbf{b}^n.
    \end{gathered}
\end{equation}

In terms of $u_\parallel$, the Lorentz factor $\Gamma$ can be written as,
\begin{equation}\label{eq:17}
    \begin{gathered}
    \Gamma = \kappa\sqrt{1+(u_\parallel^2 + 2\mu B \kappa/m)},
    \end{gathered}
\end{equation}
where $\mu = m\gamma ^2 v^2_\perp/2B$ is the relativistic magnetic moment, which for the GCA is kept constant to update $v_\perp$.

Substituting equation \ref{eq:16} into equation \ref{eq:15} for $u_\parallel$ and inverting equation $\ref{eq:17}$, we obtain:
\begin{equation}\label{eq:18}
    \Gamma^n_{n+1/2} = \max\left\{\frac{-k_2 \pm \sqrt{k_2^2 - 4k_1k_3}}{2k_1}\right\},
\end{equation}
where
\begin{equation}\label{eq:19}
    \begin{gathered}
    k_1 = \left(\frac{\Delta t}{2}\right)^2\left[\textbf{v}_E^n\cdot(\textbf{v}_E^n\cdot\grad)\textbf{b}^n\right] ^2- \\ \left(\frac{1}{\kappa^n}\right)^2\left[1-\frac{\Delta t}{2}\textbf{v}_E^n\cdot(\textbf{b}^n\cdot\grad)\textbf{b}^n\right] ^2,
    \end{gathered}
\end{equation}
\begin{equation}\label{eq:20}
    k_2 = u_\parallel' \Delta t \textbf{v}_E^n \cdot(\textbf{v}_E^n\cdot\grad)\textbf{b}^n,
\end{equation}
\begin{equation}\label{eq:21}
    \begin{gathered}
    k_3 = u_\parallel'^2 + \left(1+\frac{2\mu B^n \kappa^n}{m} \right) \cdot \\ \left[1-\frac{\Delta t}{2}\textbf{v}_E^n \cdot(\textbf{b}^n \cdot\grad)\textbf{b}^n\right] ^2.
        \end{gathered}
\end{equation}
We also define a relevant error value,
\begin{equation}\label{eq:22}
    \begin{gathered}
    \delta^k = \textbf{R}^k-\textbf{R}^n-\frac{\Delta t}{2}\Bigg[\Bigg.u_\parallel^{n+1/2}\left(\frac{\textbf{b}^k}{\Gamma^k_{n+1/2}} + \frac{\textbf{b}^n}{\Gamma^n_{n+1/2}}\right)\\- (\textbf{v}_E^k + \textbf{v}_E^n) - (\textbf{c}^k_{n+1/2} + \textbf{c}^n_{n+1/2})\Bigg.\Bigg] .
    \end{gathered}
\end{equation}
We implement the GCA with the following procedure:
\begin{enumerate}
    \item Define an electromagnetic field that can depend on space and time, then initialize the particle's $\Gamma$, $\textbf{u}$, and $\textbf{R}$, and choose a constant for $\mu$.
    \item Equation \ref{eq:13} is implicit due to terms involving $\textbf{R}^{n+1}$ on both sides of the equation. Thus, on the $n$-th iteration, we first approximate the value for $\textbf{R}^{n+1}$ with $\textbf{R}^n$ on the right hand side and calculate the new $\textbf{R}^{n+1}$. Define this value as the new approximation for $\textbf{R}^{n+1}$.
    \item Calculate $\delta$ for this approximation with equation \ref{eq:22}.
    \item Using the approximation, recalculate $\textbf{R}^{n+1}$. Define this as the new approximation.
    \item Repeat steps 3-4 until $\delta^k$ is less than a certain value (we typically set $~10^{-7}$). The final $\textbf{R}^{n+1}$ is the next position step. Update relevant parameters for the next timestep.
\end{enumerate}

Next, we will test the implementation of our numerical schemes.

\section{Simulations}
\subsection{$\mathbf{E} \times \mathbf{B}$-Drift}
\begin{figure}
    \centering
    \begin{subfigure}[b]{0.27\textwidth}
        \centering
        \includegraphics[width=\textwidth, scale = 0.15]{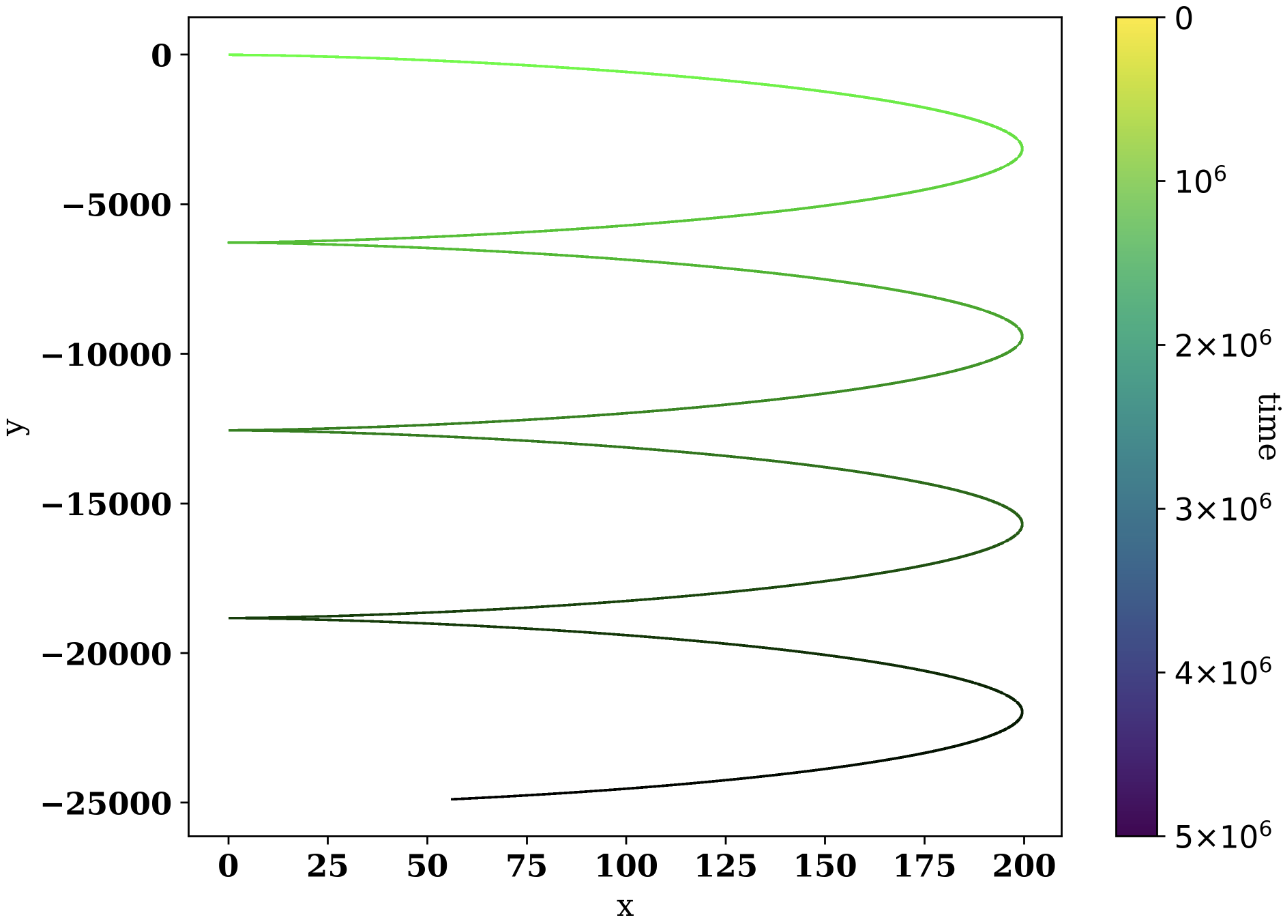}
        \caption{$xy$ plane}
        \label{1a}
    \end{subfigure}
    \hfill
    \begin{subfigure}[b]{0.211\textwidth}
        \centering
        \includegraphics[width=\textwidth, scale = 0.16]{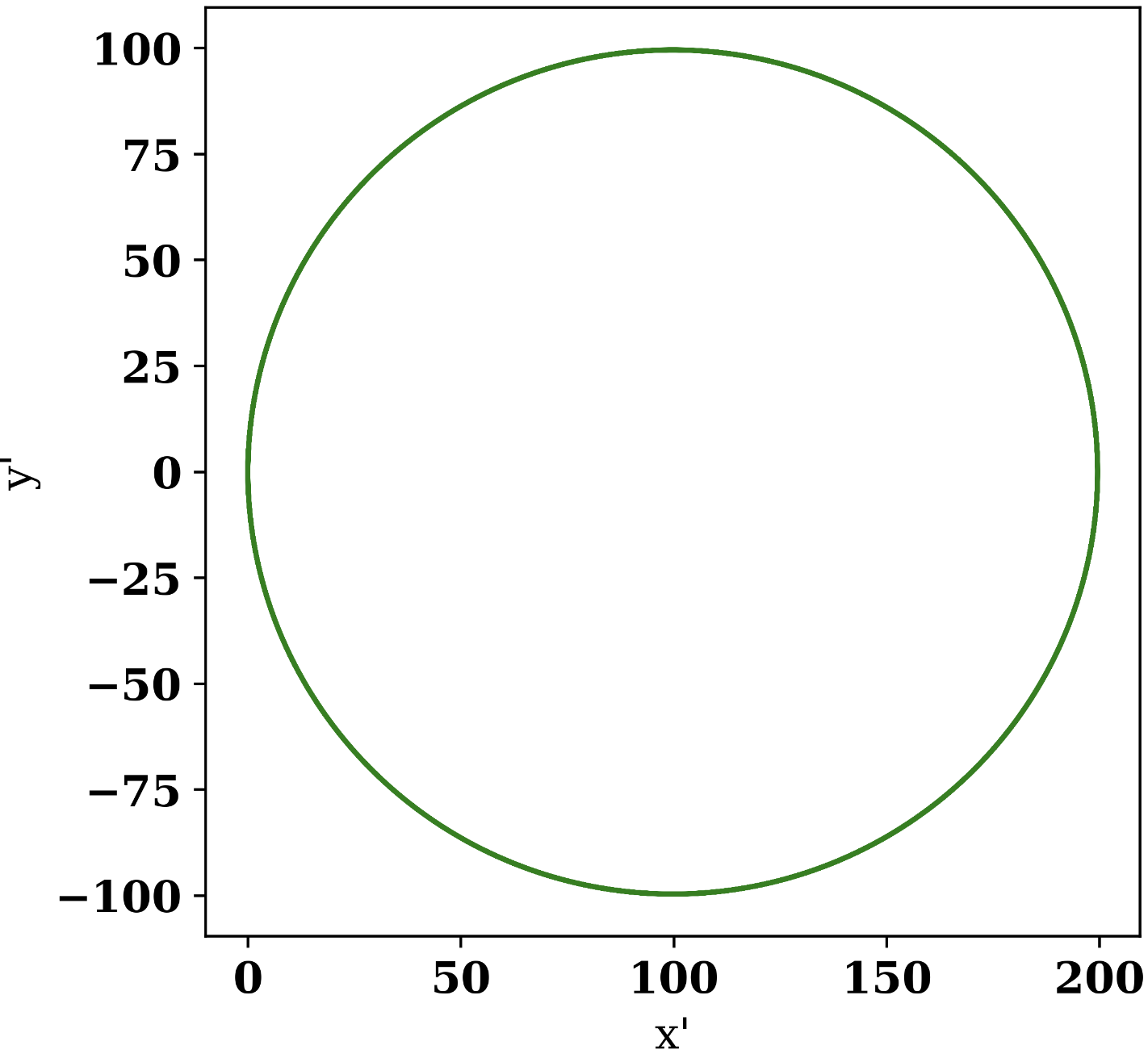}
        \caption{$x'y'$ plane}
        \label{1b}
    \end{subfigure}
    \caption{E-cross-B drift simulation using the Boris method in transformed and untransformed coordinates.}
    \label{fig:1}
    \end{figure}

    %frame transformation --> left with gyration
    The $\mathbf{E} \times \mathbf{B}$-drift describes particle motion under perpendicular magnetic and electric fields. We initialize the particle at $\textbf{x} = \langle 0, 0, 0\rangle$ with $\textbf{u} = \langle 0, 0, 0\rangle$. Following \cite{Ripperda2018}, we set $\textbf{B} = \langle B_0, 0, 0\rangle$ and $\textbf{E} = \langle0, 0, E_0\rangle$, where we choose $B_0 = 1$ and $E_0 = B_0(1-5\cdot10^{-3})$. This results in a drift Lorentz factor $\kappa = 1/\sqrt{1-v_E^2} = 10$. We set $\Delta t = 5\cdot10^{-3}$ for $5\cdot 10^6$ iterations. Figure \ref{1a} shows the results for the simulation in the $xy$ plane, where the color bar correlates the time progression with a color gradient.
    
    A frame transformation allows us to obtain a simple gyration in the $xy$ plane from the $\mathbf{E} \times \mathbf{B}$-drift \cite{Ripperda2018}.
    % \begin{gather}
    % x' = x, \\
    % y' = \kappa(y-v_E t),\\
    % z' = z,\\
    % v_x' = \frac{v_x}{\kappa}\frac{1}{1-v_Ev_y},\\
    % v_y' = \left(\frac{v_y}{\kappa} - v_E + \frac{\kappa v_y v_E^2}{\kappa + 1}\right)\frac{1}{1-v_Ev_y},\\
    % \gamma' = \kappa\gamma\left(1-v_Ev_y\right).
    % \end{gather}
    In this frame, $\textbf{E}' = 0$ and $\textbf{B}' = \textbf{B}/\kappa$, and thus the plot under the described coordinate transformation should result in a simple gyration with radius $R_c = m\gamma'v_\perp/qB'$, where $\gamma' = \kappa\gamma\left(1-v_Ev_y\right)$. Figure \ref{1b} shows the trajectory after coordinate transformation in the $x'y'$ plane, which is, as expected, a circle with radius $R_c$. The ability to transform the electric field and turn the trajectory into a gyration by switching frames shows the relationship between the electric and magnetic fields: the magnetic force in the Lorentz force can be transformed into an electric force and vice-versa.
    
\subsection{Magnetic Null}
A magnetic null refers to a region in space with a magnetic field of strength zero, representative of e.g., a reconnection layer around a neutron star powering high-energy and coherent radio emission \cite{Hakobyan2022}. A magnetic field with a null region is given by \begin{equation}\label{eq:24} \textbf{B} = B_0 \langle y/L_0, x/L_0, 0 \rangle.\end{equation} Here we set $B_0 = 1$ and $L_0 = 1$.
We initialize 40 particles of charge-to-mass ratio $q/m = 1$ uniformly spread on a unit circle on the $xy$ plane centered at the origin. They all have velocity vectors of magnitude $10^{-1}$ and pointing towards the magnetic null. We run simulations with both the Boris and GCA pushers with $\Delta t = 5\cdot10^{-2}$ and 1000 iterations. 
\begin{figure}
    \centering
    \begin{subfigure}[b]{0.238\textwidth}
        \centering
        \includegraphics[width=\textwidth, scale = 0.15]{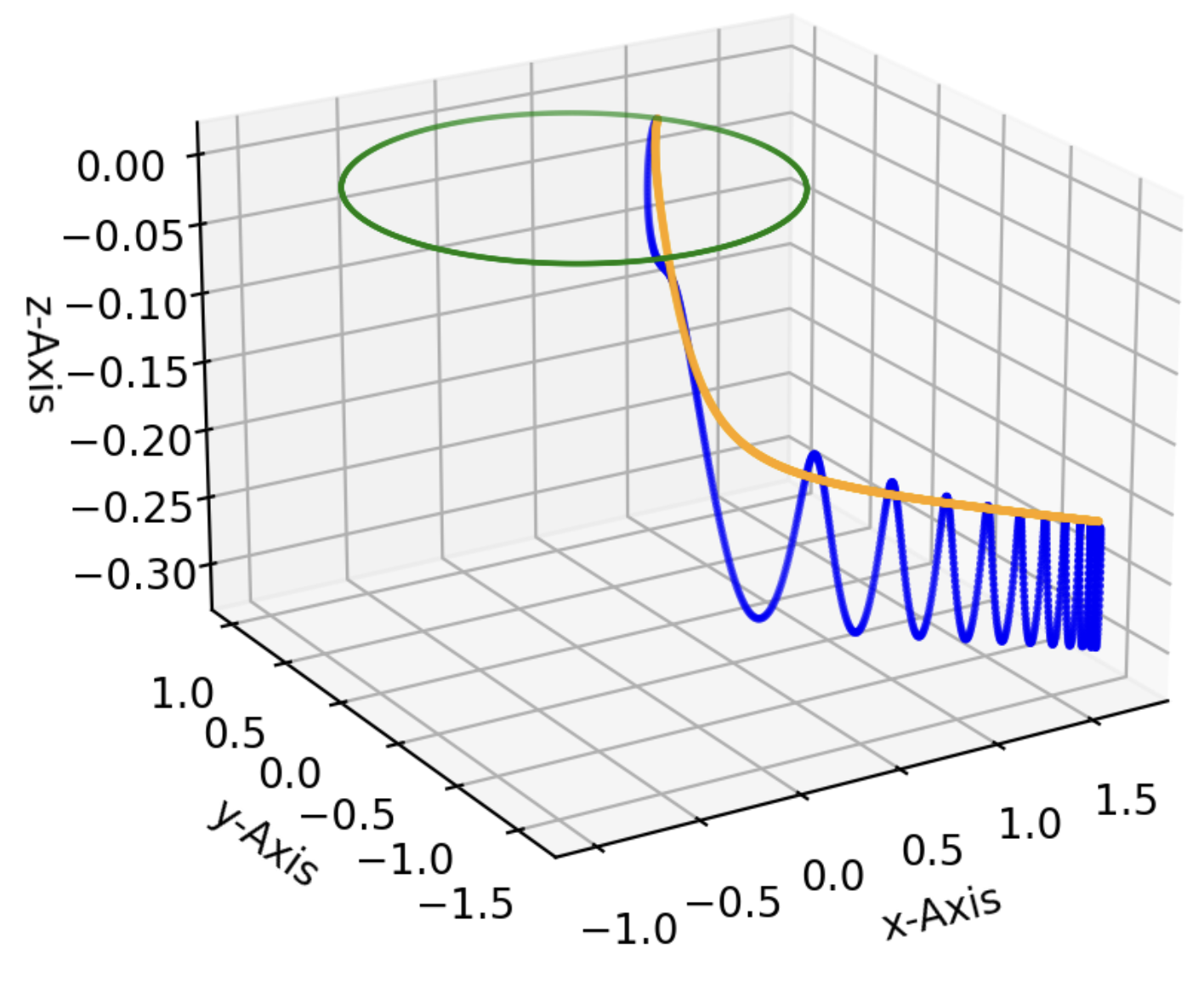}
        % \caption{Particle 4}
        \caption{}
        \label{2a}
    \end{subfigure}
    \hfill
    \begin{subfigure}[b]{0.238\textwidth}
        \centering
        \includegraphics[width=\textwidth, scale = 0.16]{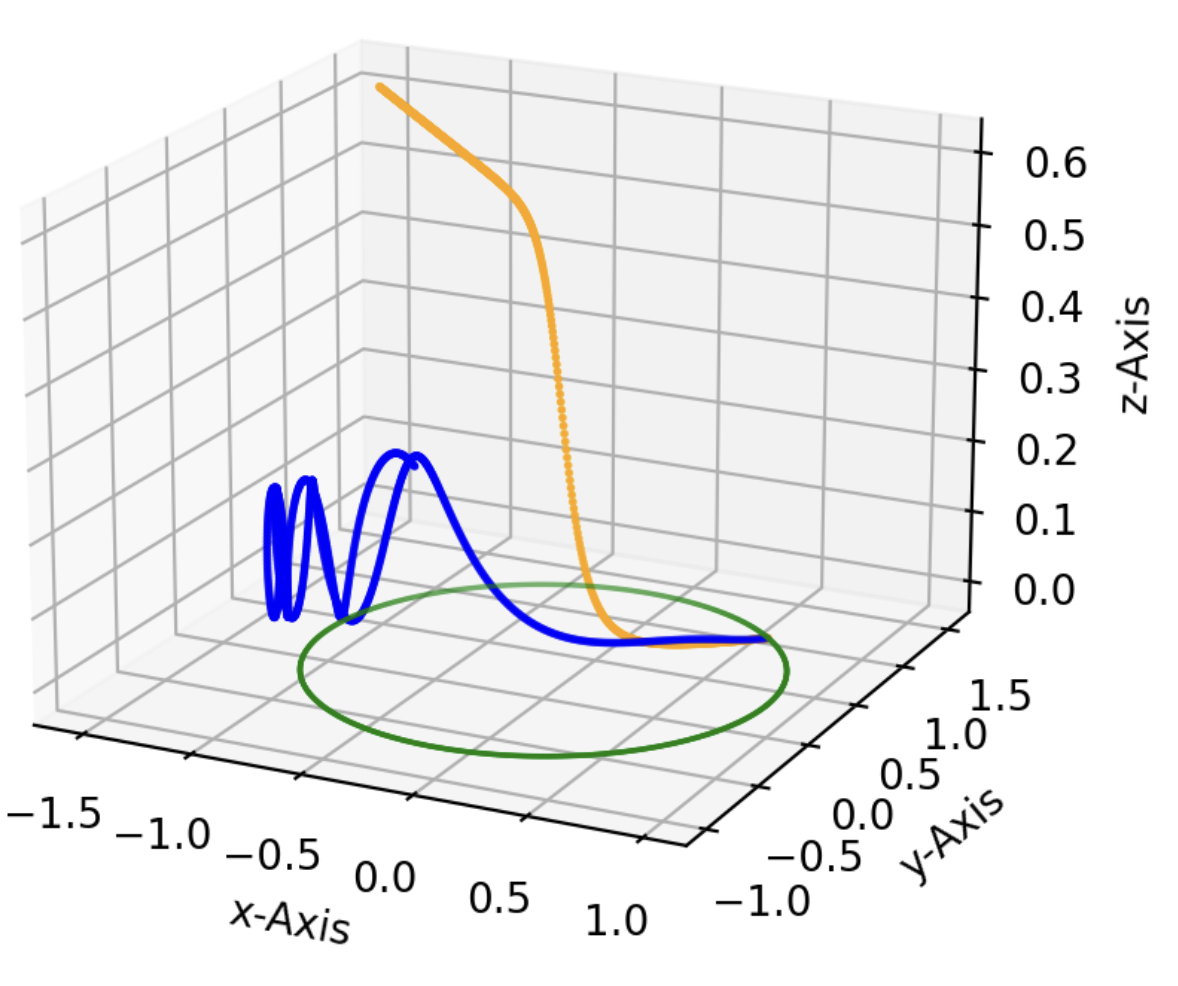}
        % \caption{Particle 5}
        \caption{}
        \label{2b}
    \end{subfigure}
    % \hfill
    % \begin{subfigure}[b]{0.23\textwidth}
    %     \centering
    %     \includegraphics[width=\textwidth, scale = 0.15]{mag_null_p20.png}
    %     \caption{Particle 20}
    %     \label{2c}
    % \end{subfigure}
    % \hfill
    % \begin{subfigure}[b]{0.23\textwidth}
    %     \centering
    %     \includegraphics[width=\textwidth, scale = 0.15]{mag_null_p33.png}
    %     \caption{Particle 33}
    %     \label{2d}
    % \end{subfigure}
    \caption{Two selected particles out of the 40 total simulated, initialized on the green unit circle. The blue curve indicates the Boris trajectory and the orange line is the GCA motion.}
    \label{fig:2}
\end{figure}

Figure \ref{fig:2} shows two selected particles out of all simulated ones. Most of the guiding center trajectories follow the Boris trajectories, as shown by subfigure \ref{2a}. They typically have a displacement from the true center of the Boris trajectory, which is expected, as the initial condition for GCA differs from the exact center of the gyromotion in the Boris pusher by a vertical displacement. Some guiding center trajectories, such as that of subfigure \ref{2b}, diverge from the Boris trajectory due to this mismatch on initial condition and the resulting orbit sampling a different magnetic field, a particle going through the magnetic null (where the GCA fails because the Larmor radius goes to infinity), or a drift that is not captured by our reduced set of GCA equations\footnote{In this case we could switch to the Boris method dynamically, instead of adding the ignored terms \cite{Bacchini2020}.}. %Figure \ref{2c} shows a particle mirroring in its motion, which is not captured in the GCA due to the approximation made in equation \ref{eq: 10}

\subsection{Magnetic Dipole, Quadrupole, and Quadrudipole}
    Magnetic dipoles are one of the most prevalent magnetic fields: neodymium magnets, current-carrying loops, and stellar objects are all sources of dipolar fields. The dipole vector potential in spherical coordinates ($r,\theta,\phi$) is given by \begin{equation}\label{eq:25} \textbf{A}_{\text{dipole}} = \left(\frac{\mu_0}{4\pi}\right)\frac{\mu_{\text{dipole}}\cdot\sin{\theta}}{r^2}\hat{\boldsymbol{\phi}},\end{equation} where $\mu_0$ is the vacuum permeability and $\mu_{\text{dipole}}$ is the dipole moment. $\hat{\textbf{e}}$ represents the unit vector in the direction of an arbitrary vector $\textbf{e}$. The magnetic field vector can be found by taking the curl of the vector potential,
    \begin{equation}\label{eq:26} \textbf{B}_{\text{dipole}} = \grad \times \textbf{A}_{\text{dipole}} = \frac{M_{\text{dipole}}}{r^3}(2\cos(\theta)\hat{\textbf{r}} + \sin{\theta}\hat{\boldsymbol{\theta}}),\end{equation} where $M_{\text{dipole}} = \mu_0\mu_{\text{dipole}}/4\pi$. In Cartesian coordinates,
    \begin{equation}\textbf{B}_{\text{dipole}} = \frac{M_{\text{dipole}}}{(x^2+y^2+z^2)^{5/2}} [3zx\hat{\textbf{x}} + 3zy\hat{\textbf{y}} + (2z^2-x^2-y^2)\hat{\textbf{z}}].\end{equation}
    
    % Standard spherical vectors can be transformed into the Cartesian plane with the following transformations:
    % \begin{gather}
    % r = \sqrt{x^2+y^2+z^2},\\
    % \cos{\theta} = \frac{z}{r},\\
    % \tan{\phi} = \frac{y}{x},
    % \end{gather}
    % and the unit vectors are represented by,
    % \begin{gather}
    % \hat{\textbf{r}} = \sin{\theta}\cos{\phi} \hat{\textbf{x}} + \sin{\theta} \sin{\phi} \hat{\textbf{y}} + \cos{\theta} \hat{\textbf{z}}, \\
    % \hat{\boldsymbol{ \theta }} = \cos{\theta}\cos{\phi} \hat{\textbf{x}} + \cos{\theta} \sin{\phi} \hat{\textbf{y}} - \sin{\theta}\hat{\textbf{z}}, \\
    % \hat{\boldsymbol{\phi}} = - \sin{\phi} \hat{\textbf{x}} + \cos{\phi}\hat{\textbf{y}}.
    % \end{gather}
    % Thus, in Cartesian coordinates, the magnetic dipole field is given by \begin{equation}\label{eq:26}\textbf{B}_{\text{dipole}} = \frac{M_{\text{dipole}}}{(x^2+y^2+z^2)^{5/2}} [3zx\hat{x} + 3zy\hat{y} + (2z^2-x^2-y^2)\hat{z}].\end{equation}
    \begin{figure}
        \centering
        \includegraphics[width=0.40\textwidth, scale = 0.15]{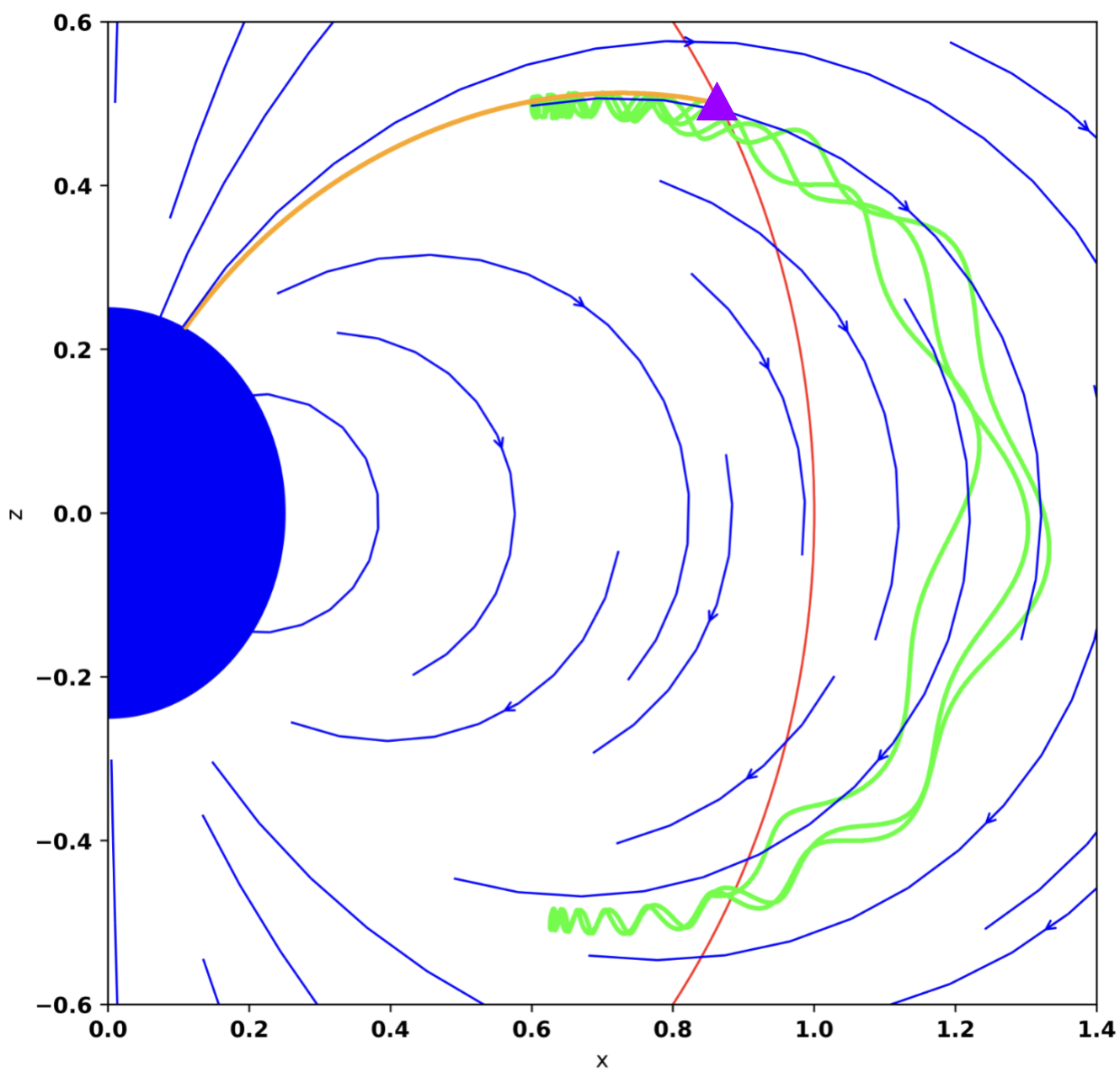}
        \caption{A single particle is simulated with the GCA (orange) and the Boris method (green). The particle is initialized $4r^*$ away from the origin in the negative radial direction. The purple triangle marks the initial position of the particle.}
        \label{fig:3}
    \end{figure}
    \begin{figure*}
    \centering
    \begin{subfigure}[b]{0.32\textwidth}
        \centering
        {\includegraphics[width=\textwidth, scale = 0.15]{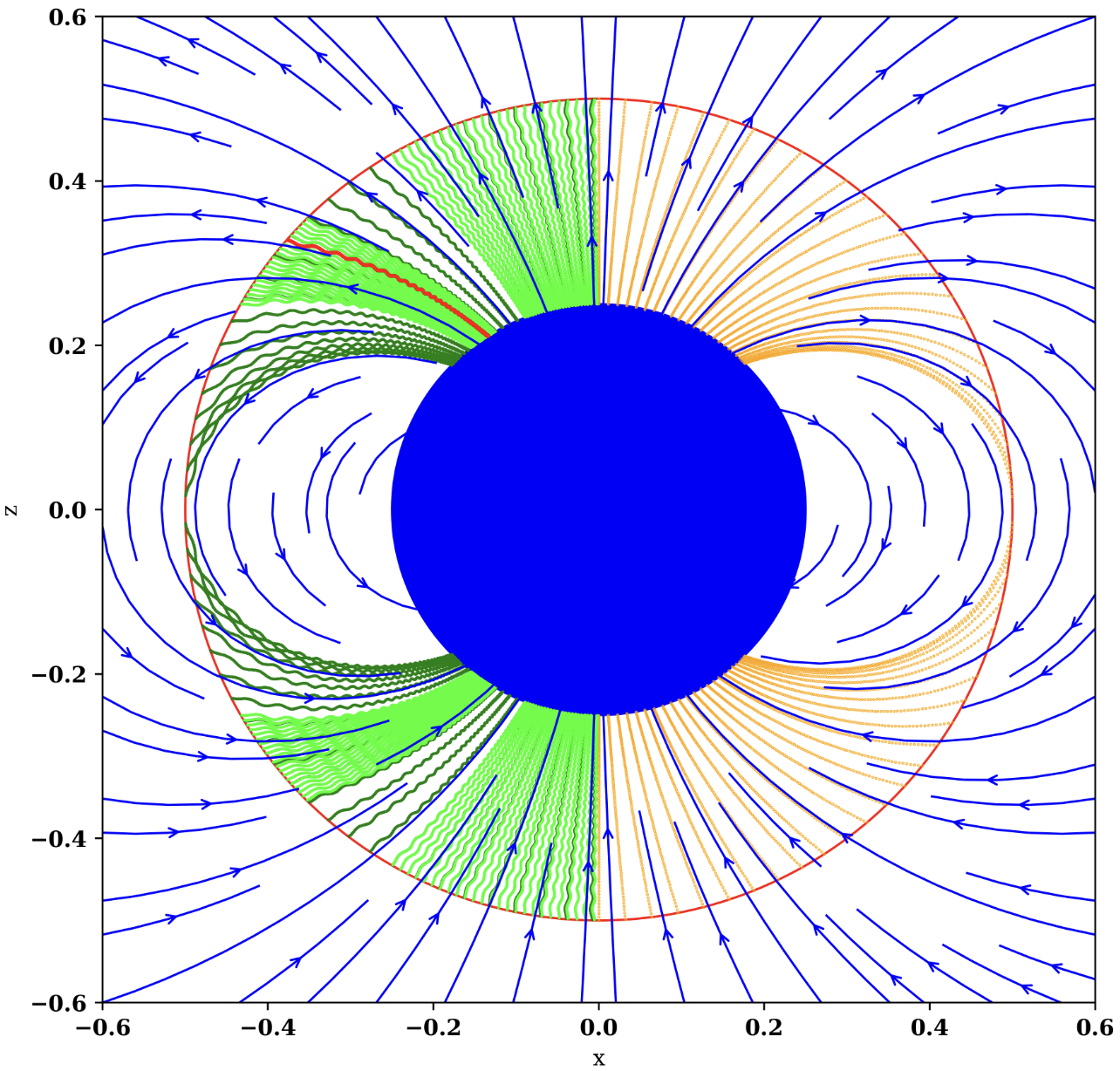}}
        \caption{Magnetic Dipole}
        \label{4a}
    \end{subfigure}
    \hfill
    \begin{subfigure}[b]{0.32\textwidth}
        \centering
        {\includegraphics[width=\textwidth, scale = 0.15]{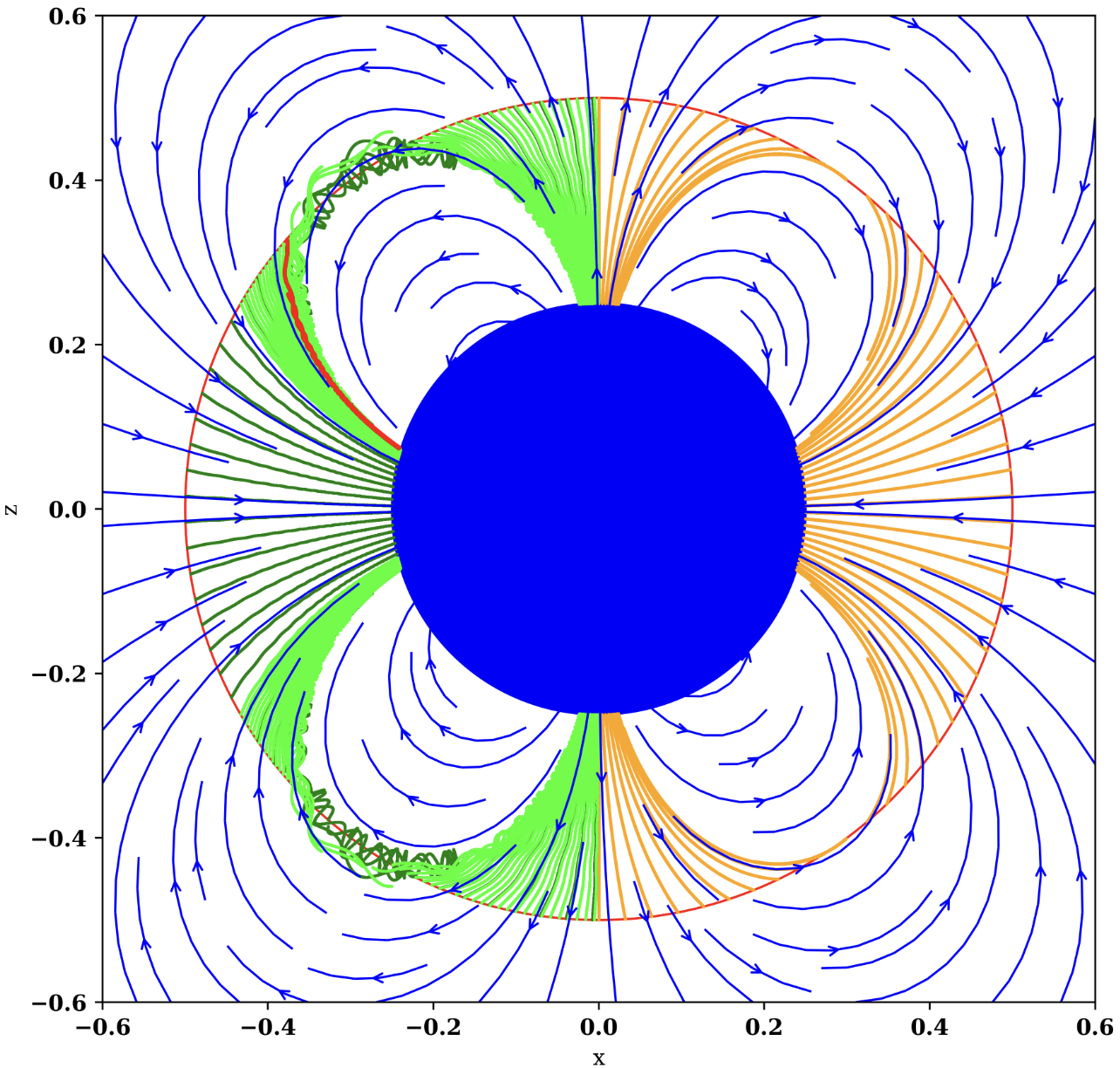}}
        \caption{Magnetic Quadrupole}
        \label{4b}
    \end{subfigure}
    \hfill
    \begin{subfigure}[b]{0.315\textwidth}
        \centering
        {\includegraphics[width=\textwidth, scale = 0.15]{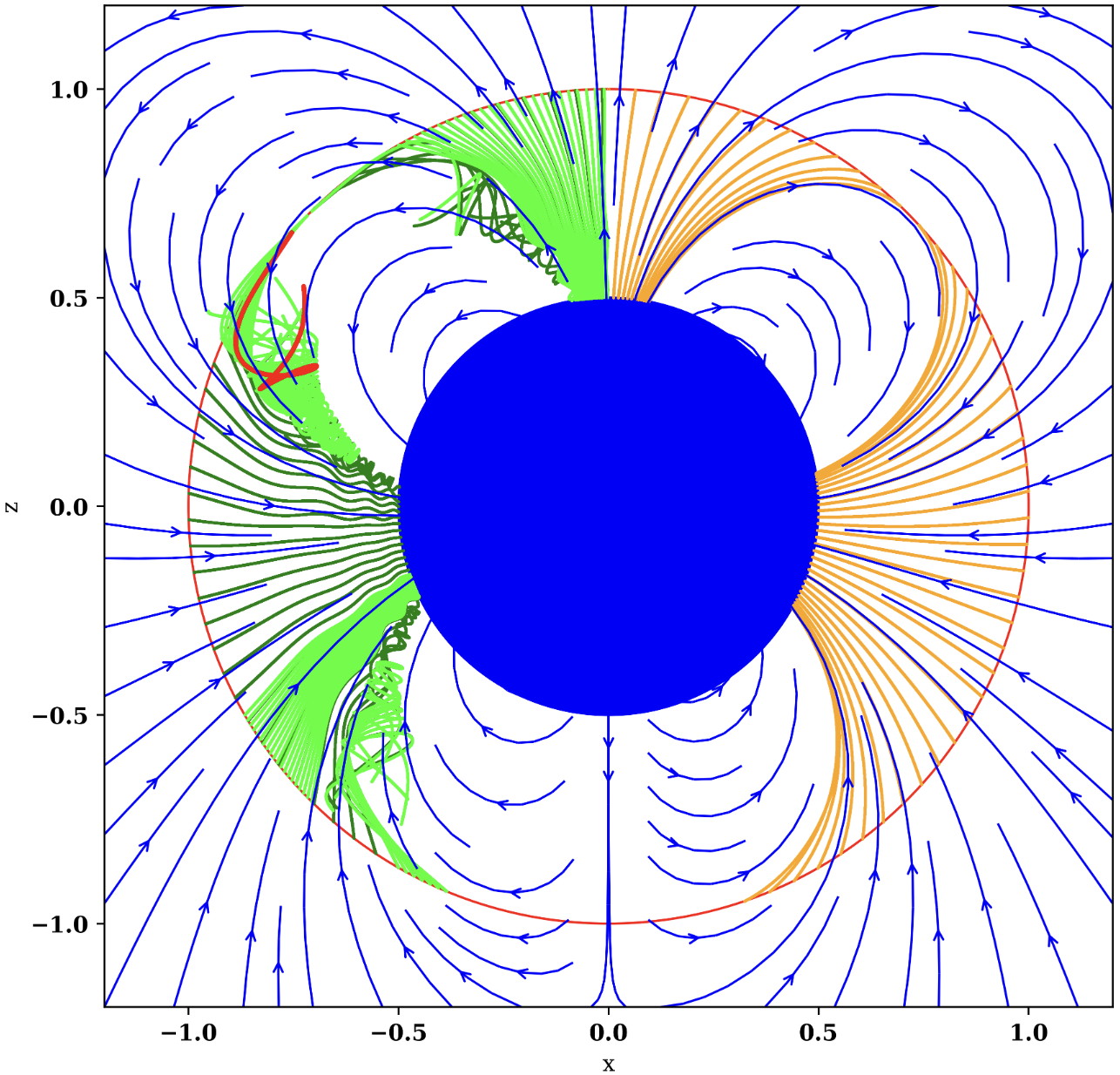}}
        \caption{Magnetic Quadrudipole}
        \label{4c}
    \end{subfigure}
    \caption{Particles simulated in magnetic dipolar, quadrupolar, and quadrudipolar fields. The solid blue circle represents the neutron star and its magnetic field lines are approximated by the blue stream lines surrounding it. The left half displays the Boris simulations, where 100 particles are initialized such that they are uniformly distributed on the red circle, represented by green trajectory lines. 180 particles are initialized in the polar caps and the separatrix, represented by lime trajectory lines. The right half shows the GCA solutions, marked by orange lines. The selected particle's trajectory is marked in red.}
    \label{fig:4}
    \end{figure*}
    
    A quadrupole is an extension of the dipole (e.g. Helmholtz Coil). Its magnetic field in polar coordinates is \cite{gu2019}
    \begin{equation}
        \begin{gathered}
        \textbf{B}_{\text{quad}} = \frac{M_{\text{quad}}}{r^4}[-(3\sin^2{\theta}\cos^2{\phi}-1) \hat{\textbf{r}} + \\ (\sin{2\theta}\cos^2{\phi} )\hat{\boldsymbol{\theta}} - (\sin{\theta}\sin{2\phi} )\hat{\boldsymbol{\phi}}],
        \end{gathered}
    \end{equation}
    where we define $M_{\text{quad}} = A\mu_{\text{quad}}$ for some constant $A$ and quadrupole moment $\mu_{\text{quad}}$. We similarly transform the quadrupolar magnetic field to Cartesian coordinates.
    
    The quadrudipolar magnetic field is a linear combination of a dipolar and quadrupolar field, so $\textbf{B}_{\text{quadrudipole}} = c_1\textbf{B}_{\text{dipole}} + c_2\textbf{B}_{\text{quad}}$,
    where we define the dipole-quadrupole ratio $z = c_1/c_2$ for some constants $c_1$ and $c_2$.
    
    We first initialize a total of $280$ particles with $q/m = 1$ in a dipolar field with $M_{\text{dipole}} = 40$, associated to a neutron star with radius $r^*$. The particles are initialized on a circle centered at the origin with radius $2r^*$. All particles possess an initial velocity parallel to the magnetic field, with a component pointing in the negative radial direction. 100 particles are uniformly distributed in their starting angular locations from $0$ to $2\pi$. 180 particles are specifically initialized from the intervals $(\pi/3, 2\pi/3), (\pi/6, \pi/4),$ and $(3\pi/4, 5\pi/6)$, as well as those obtained by reflecting these intervals across $z = 0$. We initialize more particles in these regions to represent particles that are created via pair production near the polar cap and the separatrix \cite{Philippov2015} \cite{Bransgrove2022}. We fol low the same procedure for the magnetic quadrupole and quadrudipole, choosing $M_{\text{quad}} = 1$ and $z = 1/100$ to balance the strength of the two combined fields so one does not dominate the other.

    A particle will sometimes perform ``mirror" motion in a magnetic dipole (or quadrupole \& quadrudipole) similar to charged particles in a magnetic bottle. However, as mentioned, the GCA equations we implement do not include the mirror effect. Thus, as shown in figure \ref{fig:3}, while the GCA solution depicts the particle falling straight into the star, the accurate Boris solution depicts the particle performing mirror motion around the star.
    
    Figure \ref{fig:4} shows the results for our simulations. The left half of each subfigure shows the Boris scheme simulations, while the right half shows the GCA, where we only display the uniformly initialized particles. The GCA and Boris solutions are very similar, only differentiated by the Boris solution's gyration perpendicular to the field (i.e., into the plane of the page). The GCA, while still retaining the gyration information by updating $v_\perp$, is less computationally expensive. Moreover, the typical ratio $m/q$ of an electron is on the order of $10^{-12}$, and the magnetic field strength near a neutron star can reach up to $10^9$ T. Thus, even if the particle travels at relativistic speeds, the gyroradius will be negligible such that the GCA approximation is superior to the full Boris solution for achievable time steps \cite{Philippov2020}.

    The particle trajectories for the three fields vary significantly. We mark the trail of a selected particle with red. While the initial positions of this particle are identical, the final locations vary greatly. We see that while the particle lands on the star for dipolar and quadrupolar fields, it mirrors and recedes from the star for the quadrudipole. Specifically, the final positions for the magnetic dipole and quadrupole varies by a $\Delta \theta \approx 0.288$, resulting in a physical separation of $\Delta s = 0.288 r^*$. Particles reach different positions on the star, meaning that if they have bombarded the star and potentially create a hot spot, they have sampled a different magnetic field in the dipole, quadrupole or quadrudipole. In reality this would mean that the particles encountered different electric fields and hence would have accelerated to different energies once reaching the star, potentially changing the interpretation of the hot spot. Our understanding of neutron stars may change if we consider more realistic magnetic field geometries rather than a pure magnetic dipole \cite{Gralla2016} \cite{Gralla2017}.
    
\section{Conclusion}

This research investigated the accuracy of the Boris and GCA method on simulating charged particles under various magnetic field configurations in neutron star magnetospheres. We first tested the accuracy of the two methods and then proceeded to more realistic dipolar, quadrupolar, and quadrudipolar fields. These simulations provide a potentially more realistic model of particle motion near neutron stars, which can assist the NICER mission to characterize radiation and composition of neutron stars. 

In the future, we will couple the Boris method and GCA, switching from the Boris scheme to the GCA when the Larmor radius is sufficiently small, and vice versa \cite{Bacchini2020}. We will then be able to retain nearly all information on the particle motion while significantly reducing the computation intensity. Our algorithm can further be improved by including pair production and radiative losses due to synchrotron and inverse-Compton emission, potentially driving a particle to the GCA limit \cite{Bransgrove2022} \cite{Hakobyan2022}. Finally, we can solve Maxwell's equations, simulating a changing magnetic and electric field in a dynamic magnetosphere, in a full particle-in-cell cycle.

\bibliographystyle{ieeetr}
\bibliography{citations}

\end{document}